\documentclass[
    ,final            
  ]
  {aipproc}

\layoutstyle{8x11single}

\begin{document}

\title{The status of the Excited Baryon Analysis Center}


\author{B. Juli\'a-D\'{\i}az}{
  address={Department d'Estructura i Constituents de la Mat\`{e}ria
and Institut de Ci\`{e}ncies del Cosmos,\\
Universitat de Barcelona, E--08028 Barcelona, Spain}
}

\begin{abstract}
The Excited Baryon Analysis Center (EBAC), which is 
associated with the Theory Group at Jefferson Laboratory, was 
initiated in 2006. 
Its main goal is to extract and interpret properties of nucleon 
resonances (N*) from the world data of meson production 
reactions induced by pions, photons and 
electrons. We review the main 
accomplishments of the center since then and sketch its 
near future perspectives.
\end{abstract}

\maketitle

The spectrum of low-lying nucleon and $\Delta$ resonances is 
a primordial ingredient for any understanding of the 
non-perturbative domain of strong interactions. Consequently, 
a large effort has been made during the last years to 
extract properties of $N^*$ from the world data base of 
$\pi N \to \pi N$ and $\gamma(^*) N \to \pi N$ data~\cite{blreview}.
The most relevant recent advances in our knowledge of 
$N^*$ physics are due to the effort carried out mostly 
in facilities like Jefferson Lab (USA) or MAMI and 
CB-ELSA (Germany) where the probe used is electromagnetic, 
thus permitting a cleaner access to the baryon structure. 

The use of electromagnetic probes to explore the inner 
structure of baryons does not completely avoid the 
difficulties arising from the not so well-known hadronic 
pieces of the production process. Thus, it is well 
acknowledged that one needs to attain a proper understanding 
of the hadronic interactions entering in the electromagnetic 
production processes to be able to extract any useful 
information for their analysis. This is achieved by the 
construction of involved dynamical models which incorporate 
the main physics at stake, e.g. most relevant channels, 
unitarity, and which can correlate the vast amount of data 
existing for both single and double meson production reactions. 

Among the existing theoretical approaches, the one taken 
at the Excited Baryon Analysis Center (EBAC) tries to 
encompass the above by considering the following two and three 
body channels: $\gamma N$, $\pi N$, $\pi \pi N$, $\eta N$ in a 
multi-channels multi-resonances framework~\cite{msl07}. 
The starting point of the model is a set of Lagrangians 
describing the interactions between mesons (including 
the photon) ($M$ =$\gamma$, $\pi, \eta$ , $\rho, 
\omega$, $\sigma, \dots$) and baryons ($B = N, \Delta, 
N^*, \dots$). By applying a unitary transformation 
method~\cite{jklmss09-2}, an effective Hamiltonian, with an 
energy independent set of potentials, is then derived 
from the considered Lagrangian.

The meson-baryon ($MB$) scattering amplitudes are 
obtained in the following way, 
\begin{eqnarray}
 T_{\alpha,\beta}(E)  &=&  
 t_{\alpha,\beta}(E)
+ 
 t^R_{\alpha,\beta}(E) \,,
\label{eq:tmbmb}
\end{eqnarray}
where $\alpha, \beta = \gamma N, \pi N, \eta N, \pi\pi N$. The full
amplitudes, e.g. $T_{\pi N,\pi N}(E)$, $T_{\eta N,\pi N}(E)$, 
$T_{\pi N,\gamma  N}(E)$ can be directly used to, within 
the same framework, compute $\pi N \to \pi N\,,  \eta N$ and 
$\gamma N \to \pi N$, $\gamma N \to \eta N$, scattering 
observables. The non-resonant amplitude $t_{\alpha,\beta}(E)$ 
in Eq.~(\ref{eq:tmbmb}) is defined by the coupled-channels 
equations,
\begin{eqnarray}
t_{\alpha,\beta}(E)= V_{\alpha,\beta}(E)
+\sum_{\delta}
V_{\alpha,\delta}(E) \;
G_{\delta}(E)    \;
t_{\delta,\beta}(E)  \,
\label{eq:nr-tmbmb}
\end{eqnarray}
with
$
V_{\alpha,\beta}(E)= v_{\alpha,\beta}
+Z^{(E)}_{\alpha,\beta}\,, 
\label{eq:veff-mbmb}
$
where $v_{\alpha,\beta}$ are the non-resonant $MB$ potentials and
$Z^{(E)}_{\alpha,\beta}$ is due to the one-particle-exchange between
unstable $\pi\Delta, \rho N, \sigma N$ states which are the resonant
components of the $\pi\pi N$ channel.

The second term in the right-hand-side of Eq.~(\ref{eq:tmbmb}) is 
the resonant term defined by
\begin{eqnarray} 
t^R_{\alpha,\beta}(E)= \sum_{N^*_i, N^*_j}
\bar{\Gamma}_{\alpha \rightarrow N^*_i}(E) [D(E)]_{i,j}
\bar{\Gamma}_{N^*_j \rightarrow \beta}(E) \,,
\label{eq:tmbmb-r} 
\end{eqnarray}
with
\begin{eqnarray}
[D^{-1}(E)]_{i,j} = (E - M^0_{N^*_i})\delta_{i,j} - 
 \sum_{\delta}\Gamma_{N^*_i\rightarrow \delta} G_{\delta}(E)
\bar{\Gamma}_{\delta \rightarrow N^*_j}(E) \,.
\label{eq:nstar-g}
\end{eqnarray}
where $M_{N^*}^0$ is the bare mass of the resonant state $N^*$.
The dressed vertex interactions in Eq.~(\ref{eq:tmbmb-r}) and
Eq.~(\ref{eq:nstar-g}) are (defining 
$\Gamma_{\alpha\rightarrow N^*}=\Gamma^\dagger_{N^* \rightarrow \alpha}$)
\begin{eqnarray}
\bar{\Gamma}_{\alpha \rightarrow N^*}(E)  &=&  
{ \Gamma_{\alpha \rightarrow N^*}} + \sum_{\delta}
t_{\alpha,\delta}(E) 
G_{\delta}(E)
\Gamma_{\delta \rightarrow N^*}\,, 
\label{eq:mb-nstar} \\
\bar{\Gamma}_{N^* \rightarrow \alpha}(E)
 &=&  \Gamma_{N^* \rightarrow \alpha} +
\sum_{\delta} \Gamma_{N^*\rightarrow \delta}
G_{\delta }(E)t_{\delta,\alpha}(E) \,. 
\label{eq:nstar-mb}
\end{eqnarray}

\begin{figure}[t!]
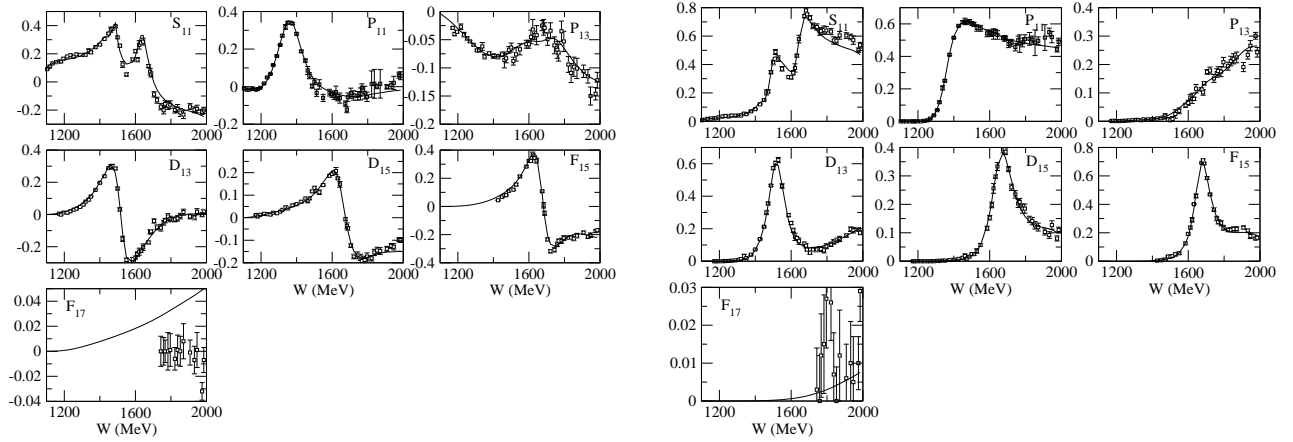

\vspace{20pt}
\includegraphics[width=8.1cm]{fig4.eps}
\hspace{0.5cm}
\includegraphics[width=8.1cm]{fig5.eps}
\caption{Real (left) and imaginary (right) parts of the calculated 
$\pi N$ partial wave amplitudes (Eq.~(\ref{eq:tmbmb})) of isospin 
$T=1/2$  are compared with the energy independent solutions of 
Ref.~\protect\cite{said06}.
\label{figrealiso1}}
\end{figure}

\section{Developments in hadronic reactions}

Within the framework sketched above we built a hadronic 
model, JLMS, which was aimed at properly describing the 
$\pi N \to \pi N$ and $\pi N \to \eta N$ experimental 
data in the energy range relevant for $N^*$ physics, 
1 GeV $<W_{\rm c.m.}<2$ GeV~\cite{jlms07}. The model was 
built by performing extensive $\chi^2$ minimizations 
to the experimental data collected from the GWU-SAID 
database~\cite{said06} and to the partial wave amplitudes 
of the GWU-SAID group. In fig.~\ref{figrealiso1} the 
real and imaginary part of the scattering amplitudes 
for $\pi N \to \pi N$ reactions are compared to the 
GWU-SAID ones for several partial waves. The agreement 
is in most waves extremely good, with the only exception 
of the $S_{31}$ partial wave, which does not agree with 
the same quality. 

The $\pi N \to \pi N$ model predictions are given in detail 
in Ref.~\cite{jlms07}, with explicit comparisons to 
experimental data both for differential cross sections 
and polarizations as well as for the total cross sections 
predicted by the model.

\subsection{$N^*$ properties from the analytic continuation of 
the scattering amplitudes}

Extracting the properties, e.g. masses, widths, and couplings 
to $MB$ channels, of the resonances from a hadronic model is 
in general not an easy task. A proper extension of the model 
to the complex energy plane is needed so that poles of the 
$t-$matrix can be isolated and their properties extracted. 
The analytic extension of the dynamical coupled-channels model 
described above has been done in Refs.~\cite{ssl1,ssl2}. In 
table~\ref{tab1} we provide the position in the complex$-E$ 
plane of all the poles present in our $\pi N$ 
model~\cite{prl10}. First, let us note that most of the pole 
positions agree within errors with those already reported 
by the PDG~\cite{pdg}. However, similarly to the recent 
developments from the GWU-SAID group~\cite{said06}, our model 
does not find any poles corresponding to several of the $N^*$ 
states present in the PDG and rated with three or less stars. 
Also, we do not find poles in the $P_{13}$ and $P_{31}$ partial waves.

\begin{table}[bt]
\caption{The resonance pole positions $M_R$ [listed as 
$({\rm Re}~M_R, -{\rm Im}~M_R)$] extracted from the JLMS 
model in the different unphysical sheets are compared 
with the values of 3- and 4-stars nucleon resonances 
listed in the PDG~\cite{pdg}. The notation indicating 
their locations on the Riemann surface are explained 
in the text. ``---" for $P_{33}(1600)$, $P_{13}$ and 
$P_{31}$ indicates that no resonance pole has  been 
found in the considered complex energy region, 
Re$(E)\leq 2000$ MeV and $-$Im$(E)\leq 250$ MeV. 
All masses are in MeV.
\label{tab1}}
\begin{tabular}{ccccl}      
         &$M^0_{N^*}$&$M_R$ & Location      & PDG   \\
\hline
$S_{11}$ &1800       &(1540,   191)&$(uuuupp)$ &(1490 - 1530, \, 45 -   125)\\
         &1880       &(1642, \, 41)&$(uuuupp)$ &(1640 - 1670, \, 75 - \, 90)\\
$P_{11}$ &1763       &(1357, \, 76)&$(upuupp)$ &(1350 - 1380, \, 80 -   110)\\
         &1763       &(1364,   105)&$(upuppp)$ &                            \\
         &1763       &(1820,   248)&$(uuuuup)$ &(1670 - 1770, \, 40 -   190)\\
$P_{13}$ &1711       &\multicolumn{2}{c}{---}  &(1660 - 1690, \, 57 -   138)\\
$D_{13}$ &1899       &(1521, \, 58)&$(uuuupp)$ &(1505 - 1515, \, 52 - \, 60)\\
$D_{15}$ &1898       &(1654, \, 77)&$(uuuupp)$ &(1655 - 1665, \, 62 - \, 75)\\
$F_{15}$ &2187       &(1674, \, 53)&$(uuuupp)$ &(1665 - 1680, \, 55 - \, 68)\\
$S_{31}$ &1850       &(1563, \, 95)&$(u-uup-)$ &(1590 - 1610, \, 57 - \, 60)\\
$P_{31}$ &1900       &\multicolumn{2}{c}{---}  &(1830 - 1880,   100 -   250)\\
$P_{33}$ &1391       &(1211, \, 50)&$(u-ppp-)$ &(1209 - 1211,
\, 49 - \, 51)\\
        &1600       & \multicolumn{2}{c}{---} &(1500 - 1700, 200 -  400)\\
$D_{33}$ &1976       &(1604,   106)&$(u-uup-)$ &(1620 - 1680, \, 80 -   120) \\
$F_{35}$ &2162       &(1738,   110)&$(u-uuu-)$ &(1825 - 1835,   132 -   150)\\
         &2162       &(1928,   165)&$(u-uuu-)$ &                            \\
$F_{37}$ &2138       &(1858,   100)&$(u-uuu-)$ &(1870 - 1890,   110 -   130)
\end{tabular}
\end{table}

\subsection{A new perspective on $P_{11}$ nucleon resonances}

As described in detail in Ref.~\cite{prl10} the determination 
of resonance poles in the $P_{11}$ partial wave has been difficult 
since the discovery of the Roper resonance in 1964. In our 
model we find two poles near the PDG value  
$({\rm Re}~M_R, -{\rm Im}~M_R) =$ (1350$-$1380, 80$-$110) 
corresponding to the Roper, $N^*$(1440), resonance. This 
finding is consistent with the results from the analysis 
by Cutkosky and Wang~\cite{cw90} (CMB), GWU/VPI~\cite{said06} 
and  J\"{u}lich ~\cite{juelich}. 

\begin{figure}[t]
\includegraphics[width=6.cm]{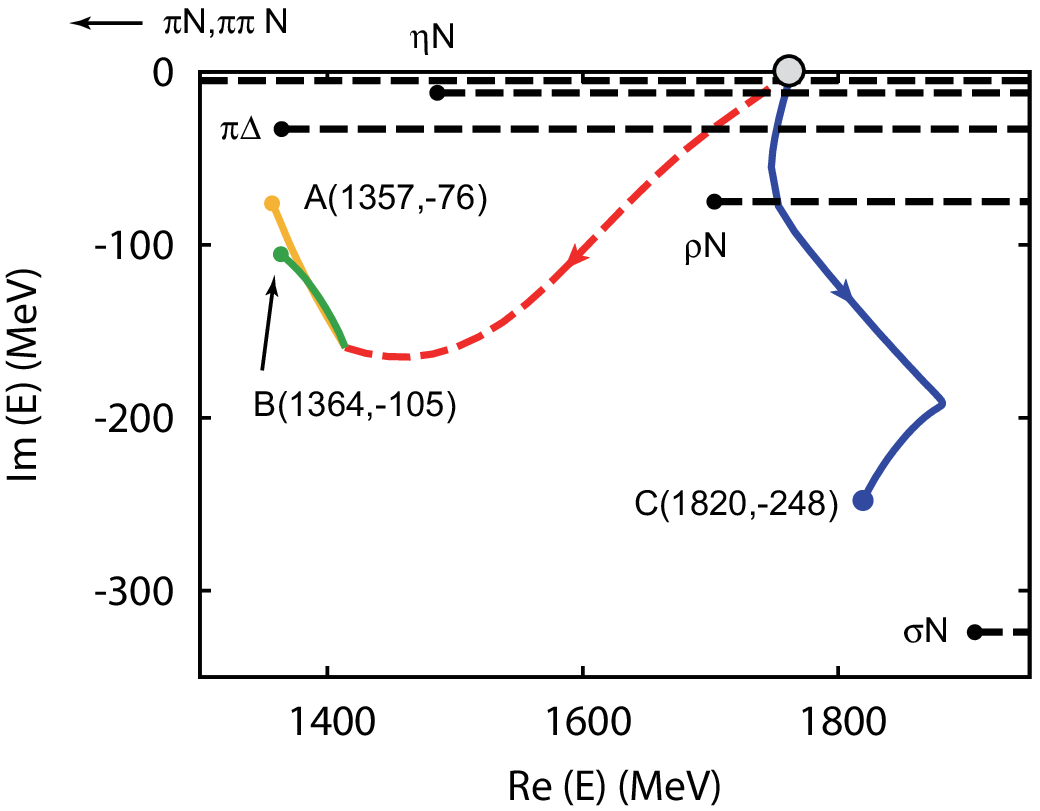}
\hspace{1cm}
\includegraphics[width=6.cm]{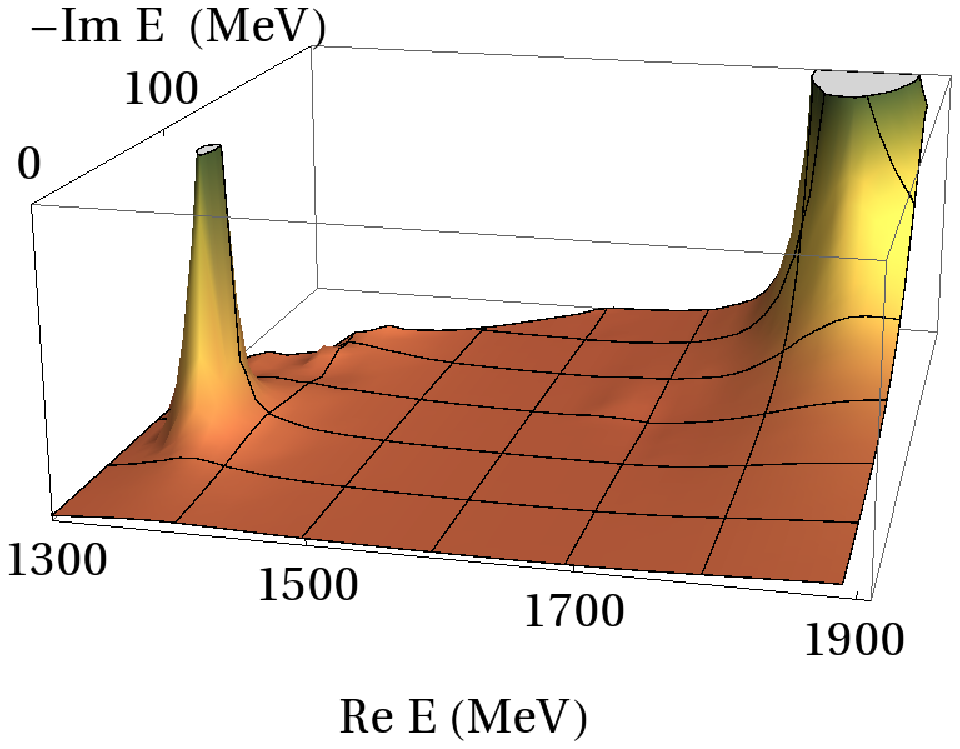}
\caption{
(left) 
Trajectories of the evolution of $P_{11}$ resonance 
poles A (1357,76), B (1364,105), and C (1820,248) from 
a bare $N^*$ with 1763 MeV, as the couplings of the 
bare $N^*$  with the meson-baryon reaction channels 
are varied from zero to the full strengths of the JLMS 
model. See text for detailed explanations. Brunch cuts for all channels
are denoted as dashed lines.
The branch points, $E_{\rm b.p.}$, for  unstable  channels are
determined by  $E_{\rm b.p.}- E_M(k)-E_B(k)- \Sigma_{MB}(k,E_{\rm b.p.})=0$ 
of the their propagators (described in the text) evaluated at the spectator 
momentum $k$=0. With the parameters~\cite{msl07} used in JLMS 
model, we find that $E_{\rm b.p.}$ (MeV) $ =( 1365.40, - 32.46), 
(1704.08, -74.98),( 1907.57, -323.62)$ for $\pi\Delta$, $\rho N$, 
and $\sigma N$, respectively. (right) 3-Dimensional depiction 
of the behavior of $\left|{\rm det}[ D(E)]\right|^{2}$ of 
the $P_{11}$ $N^*$ propagator (in arbitrary units) as a 
function of complex-$E$. 
\label{fig:p11-pole} }
\end{figure}

A higher mass pole at $(1820, 248)$ in the same partial wave,  
which is close to the $N^*(1710)$ state listed by PDG is 
also found. Moreover, within our model we find that this 
pole and the two  corresponding the Roper resonance are 
related to only one bare state. This common bare state 
can be pictured by depicting the evolution of the pole 
positions as we vary the coupling to the different inelastic 
channels, see Fig.~\ref{fig:p11-pole}. 

\subsection{Analysis of the $\pi N \to \pi \pi N$ reactions}

\begin{figure}[t]
\centering
\includegraphics[clip,width=8cm]{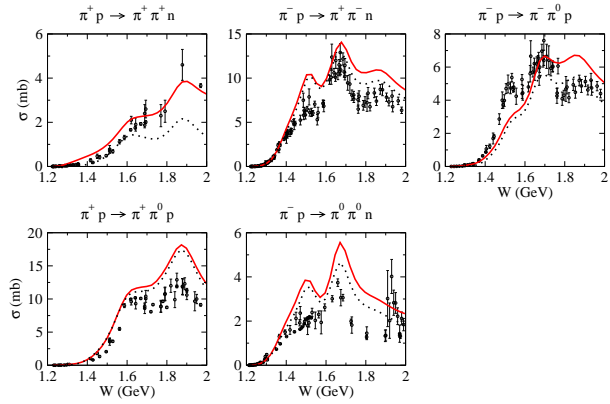}
\caption{ The total cross sections predicted
(solid curves) from the JLMS model are
compared with the experimental data. The dotted 
curves are from turning off the amplitude 
$T^{{\rm dir}}_{\pi N, \pi\pi N}$. Experimental 
data are from Ref.~\cite{datapi2pi}
}
\label{fig:jlms-pipp}
\end{figure}

The production of two or more pions provides precious 
information on the way $N^*$ couple to the higher 
meson-baryon channels. At moderately low center of 
mass energies, e.g. $W \sim 1500$ MeV, the importance 
of $\pi \pi N$ channels is already sizeable and therefore 
it is important not only to incorporate such channels 
into the framework but also to have experimental data to
 properly constrain their couplings. The analysis of the 
$\pi N \to \pi \pi N$ reaction provides a first test of 
such ingredients present in our model. This information 
would, in principle, be very useful to build the hadronic 
model. However, due to the lack of enough experimental 
data we could not profit from these data in our 
minimizations. The full details of our calculation are 
given in Ref.~\cite{kjlms09}. In figure~\ref{fig:jlms-pipp} 
we provide a comparison of the predicted total cross 
sections compared to data. In the same figure we include 
the results obtained without the inclusion of the 
direct $2-3$ mechanisms. 

\subsection{An improved model for $\pi N \to \eta N$}

The JLMS model was mostly constrained from the $\pi N \to \pi N$ 
data. Thus, it was expected that some of parameters related to the 
$\eta N$ channels were not well constrained. In Ref.~\cite{djlss08} 
we considered the full $\pi N \to \eta N$ data-base and performed 
$\chi^2$ minimizations to further constrain the parameters related 
to the $\eta N$ channel. 

\begin{figure}[thb]
\includegraphics[clip,width=7cm]{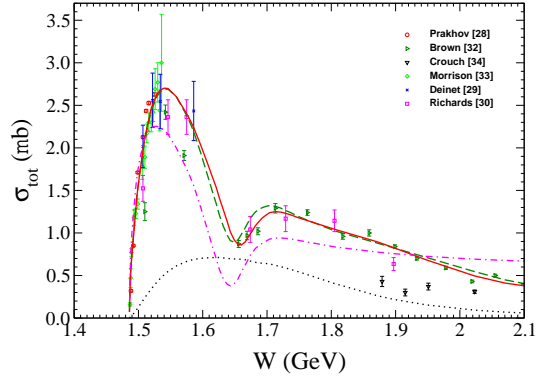}
\caption{Total cross-section for the reaction $\pi^- p \to \eta n$. 
Curves are from Ref.~\cite{jlms07} (dash-dotted), model $A$ 
(dashed), model $B$ (full), and the background contributions 
(dotted) in model $B$. Data are from \cite{dataetan}.
}
\protect\label{fig:tot}
\end{figure}

In order to determine the parameters, a data set including
294 measured differential cross-sections, coming from five 
collaborations, were fitted. The selection of data points 
allows to suppress the manifestations of inconsistencies 
among available data sets. The best model, named $B$ 
in~\cite{djlss08}, reproduces satisfactorily the data, 
with a reduced $\chi^2$ = 1.94. A detailed study of the 
reaction mechanism within the model allows to establish a 
hierarchy in the roles played by nucleon resonances. 
Actually, the dominant resonant turns out to be the 
$S_{11}(1535)$. The other resonances affecting the 
$\chi^2$ by more than 20\% when switched off, are by 
decreasing importance: $P_{11}(1440)$, $P_{13}(1720)$, 
$S_{11}(1650)$, $F_{15}(1680)$, $P_{11}(1710)$, and 
$D_{13}(1520)$. Contributions from $D_{13}(1700)$ and 
$D_{15}(1675)$ are found to be negligible. In fig.~\ref{fig:tot} 
we present a comparison of the total cross section computed
with our model with the data, more detailed information can 
be found in Ref.~\cite{djlss08}.

\section{Electromagnetic production}

As outlined in the introduction, the main aim of the EBAC 
is to analyze the extant photo and electroproduction data, 
measured mainly in Germany (MAMI and Bonn) and the USA 
(Jefferson Lab) and settle the baryon spectrum, extracting 
and interpreting the properties of the nucleon resonances.

\subsection{Single pion photoproduction}

\begin{figure}[t!]
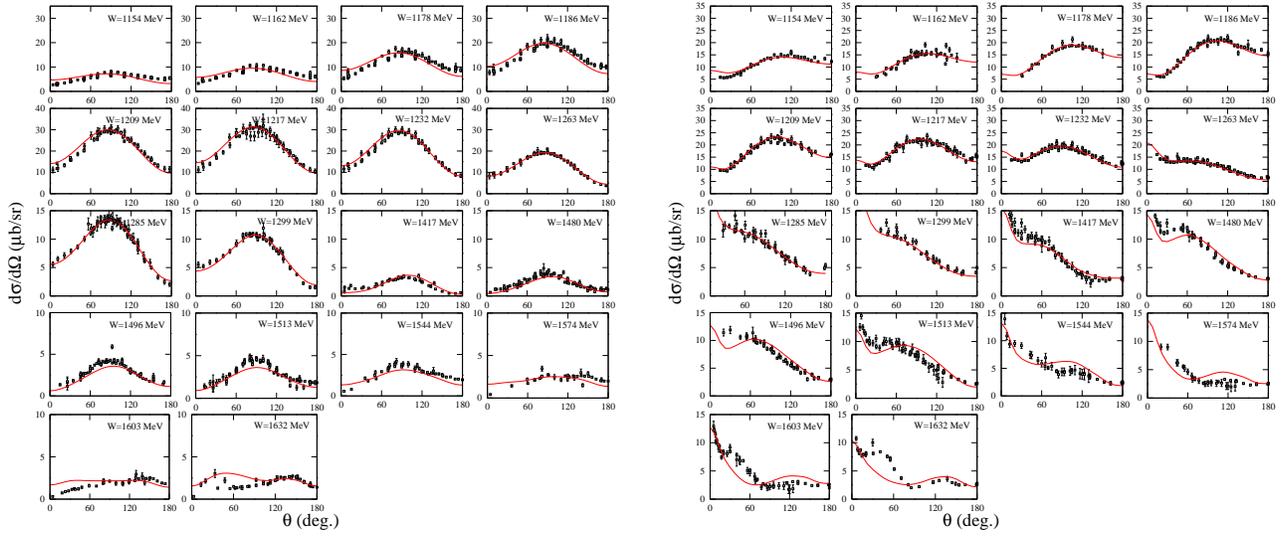

\centering
\includegraphics[width=8.1cm,angle=-00]{phot1.eps}
\hspace{0.5cm}
\includegraphics[width=8.1cm,angle=-00]{phot2.eps}
\caption{Differential cross section for $\gamma p \to \pi^0 p$ (left) and 
$\gamma p \to \pi^+ n$ (right) compared to experimental data obtained 
from Ref.~\cite{saiddb}.}
\label{fig:dcsp0m0}
\end{figure}

We have applied the dynamical coupled-channels model of 
Ref.~\cite{msl07}, outlined in the introduction, to 
investigate the pion photoproduction reactions in the 
first and second nucleon resonance region. With the 
hadronic parameters of the JLMS model of $\pi N$ scattering 
data and the non-resonant electromagnetic couplings 
taken from the previous works, we showed that the available 
data of differential cross sections and photon asymmetries of 
$\gamma N \rightarrow \pi N$ up to $W = 1.65$ GeV can be 
described to a very large extent~\cite{jlmss08}, see 
Fig.~\ref{fig:dcsp0m0}. The only free parameters in the 
$\chi^2$-fit to the photoproduction data are the bare 
$\gamma N\rightarrow N^*$ helicity amplitudes, see 
Eqs.~(\ref{eq:mb-nstar},\ref{eq:nstar-mb}). It 
is found that the coupled-channels effects can have about 
30 - 40 $\%$ effects in the $\Delta$ (1232) resonance 
region, and can drastically change the magnitudes and 
shapes of the cross sections in the second resonance 
region. We also demonstrate the importance of the 
loop-integrations in a dynamical approach. The meson cloud 
contributions to the $\gamma^* N \rightarrow N^*$ form factors 
have been predicted. For all cases, they are mainly in 
the low $Q^2$ region.  The coupled-channels effects on the 
meson cloud contributions are also found to be mainly in the 
low $Q^2$ region. 

\subsection{Double pion photoproduction}

\begin{figure}[t]
\centering
\includegraphics[clip,width=0.7\textwidth]{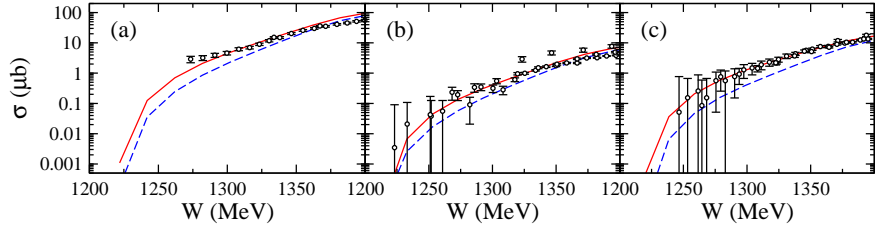}
\caption{Near threshold behavior of the total cross section for
$\gamma p \to  \pi \pi N$:
(a) $\gamma p\to\pi^+\pi^-p$,
(b) $\gamma p\to\pi^0\pi^0p$,
and (c) $\gamma p\to\pi^+\pi^0n$.
The red solid curve is the full results predicted from our current model, 
and the blue dashed curves
are the results without the $T^{{\rm dir}}_{\gamma N,\pi\pi N}$ contribution.
The data are taken from Refs.~\cite{datag2pi}.
}
\label{fig:g2ptcs-th}
\end{figure}

\begin{figure}[t]
\centering
\includegraphics[clip,width=0.4\textwidth]{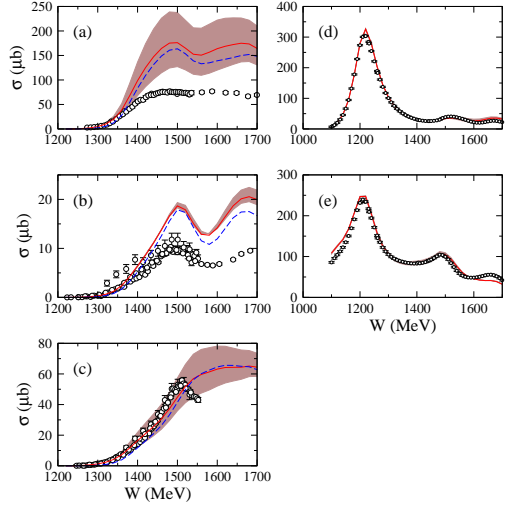}
\caption{Total cross sections of the double and single pion photoproduction
reactions up to $W=1.7$ GeV: 
(a) $\gamma p \to \pi^+\pi^-p$, 
(b) $\gamma p \to \pi^0\pi^0p$, (c) $\gamma p \to \pi^+\pi^0n$,
(d) $\gamma p \to \pi^0 p$, and (e) $\gamma p \to \pi^+n$. 
The red solid curve is the full result predicted from our current 
model, and the blue dashed curve in (a)-(c) is the result
without $T^{{\rm dir}}_{\gamma N,\pi\pi N}$ contribution. The band is generated 
by allowing a $25\%$ variation in the value of the $\pi N\Delta$ coupling
constant $g_{\pi N \Delta}$ used in the electromagnetic amplitudes. 
The data of the double and single pion photoproduction reactions 
are taken from Refs.~\cite{datag2pi}
and Refs.~\cite{saiddb}, respectively.
}
\label{fig:c3}
\end{figure}

We extended our single pion photoproduction model to predict 
the total cross sections and invariant mass distributions for 
two pion photoproduction reactions. The main aim is of course 
to be able to analyze and profit from the extensive extant 
data base. In Ref.~\cite{kjlms09-2} we preformed a detailed 
analysis of the current predictions from our framework. In 
Figs.~\ref{fig:g2ptcs-th} we present our prediction for the 
near threshold behavior of the total cross sections for 
three different production reactions. The predictions 
without any further adjusting of the parameters were not 
very satisfactory in the resonance region, thus we 
decided to explore the dependence of our predictions on 
some of the parameters of the model. In fig.~\ref{fig:c3} 
we present the effect of variations of the $\pi N\Delta$ coupling 
constant in the electromagnetic pieces. 

%
%

\begin{figure}[t]
\centering
\includegraphics[clip,width=7cm,angle=0]{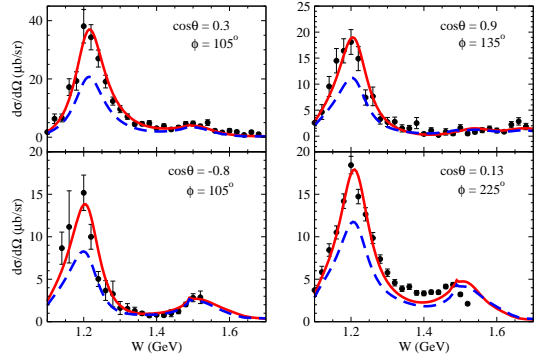}
\caption{
Coupled-channels effect on the five-fold differential cross sections 
$\Gamma_\gamma^{-1}[d\sigma^5/(d E_{e'}d\Omega_{e'}d\Omega_\pi^*)]$ 
of $p(e,e'\pi^0)p$ (upper panels) and $p(e,e'\pi^+)n$ (lower panels)
at $Q^2=0.4$ (GeV/c)$^2$.
Here $\theta\equiv\theta_\pi^*$ and $\phi\equiv\phi_\pi^*$.
The solid curves are the full results calculated with 
the bare helicity amplitudes of Fit1.
The dashed curves are the same as the solid curves but only 
the $\pi N$ loop is taken into account, see Ref.~\cite{jklmss09} 
for details. 
The data are taken from Ref.~\cite{hallb-site}}.
\label{fig:5dim}
\end{figure}

\subsection{Pion electroproduction reactions}

The framework required for the analysis of electroproduction 
reactions within the dynamical coupled channels model under 
consideration is also described in Ref.~\cite{msl07}. In 
Ref.~\cite{jklmss09} we performed the first calculations using 
such framework and the hadronic pieces of the model from 
the JLMS model. 

Electroproduction analysis have one major advantage, by 
varying the virtuality carried by the photon we can explore 
deeper into the baryon. The precise $Q^2$ evolution of the 
excitation vertexes for the different $N^*$ states 
(helicity amplitudes) is thus a subject of intensive 
study, see for instance Refs.~\cite{vlad,hwei, bruno-1}. 

The quantity relevant to our discussions is the dressed
$\gamma ^* N \rightarrow N^*$ vertex function
 defined by
\begin{eqnarray}
\bar{\Gamma}^{J}_{N^*,\lambda_\gamma\lambda_N}(q,W,Q^2)
&=&{\Gamma}^{J}_{N^*,\lambda_\gamma\lambda_N }(q,Q^2) \nonumber \\
&+ &\sum_{M'B'}
\sum_{L^{\prime}S^{\prime}}
\int k^{\prime 2}dk^{\prime }
 \bar{\Gamma}^{J}_{N^*,L'S'M'B'}(k',W) G_{M'B'}(k',W)
{\it v}^{J}_{L' S' M'B' , \lambda_\gamma\lambda_N}(k',q,Q^2)\,. \nonumber \\
& &
\label{eq:pw-v}
\end{eqnarray}
The second term of Eq.~(\ref{eq:pw-v}) is due to the mechanism 
where the non-resonant electromagnetic meson production takes 
place before the dressed $N^*$ states are formed. Similar to 
what was defined in previous works, we call this 
contribution the {\it meson cloud effect}. Let us emphasize 
that the meson cloud term in Eq.~(\ref{eq:pw-v}) is the necessary 
consequence of the unitarity conditions. How this term and the 
assumed bare $N^*$ states are interpreted is obviously model 
dependent. 

Within the one-photon exchange approximation, the differential 
cross sections of pion electroproduction can be written as 
a function of several structure functions, 
\begin{eqnarray}
\frac{d\sigma^5}{d E_{e'} d\Omega_{e'} d\Omega_\pi^*}
&=& f(\sigma_T,  \sigma_L, \sigma_{LT}, \sigma_{TT} , \sigma_{LT^\prime})
\label{eq:dcrst-em}
\end{eqnarray}
The formula for calculating $\sigma_{\alpha}$
from the amplitudes are given in Ref.~\cite{sl09}. In this first-stage 
investigation, we only considered the data of
structure functions $\sigma_\alpha$ of $p(e,e'\pi^0)p$
and $p(e,e'\pi^+)n$ up to $W=1.6$ GeV and $Q^2=1.45$ (GeV/c)$^2$.

To proceed, we need to define the bare 
$\gamma^\ast N \rightarrow N^*$ vertex functions 
$\Gamma^{J}_{N^*,\lambda_\gamma\lambda_N}(q,Q^2)$ of Eq.~(\ref{eq:pw-v}).
We parameterize these functions as
$
{\Gamma}^{J}_{N^*,\lambda_\gamma\lambda_N}(q,Q^2)
 =1/(2\pi)^{3/2} \sqrt{m_N/E_N(q)}\sqrt{q_R/|q_0|}
G_{\lambda}(N^*,Q^2) \delta_{\lambda, (\lambda_\gamma-\lambda_N)},
$ 
where $q_R$ and $q_0$ are defined by 
$M_{N^*} = q_R+E_N(q_R)$ with $N^*$ mass and
$W = q_0+E_N(q_0)$, respectively.

The only freedom in this study for analyzing the electromagnetic 
meson production reactions is the electromagnetic coupling 
parameters of the model. If the parameters
listed in Ref.~\cite{msl07} are used to calculate the 
non-resonant interaction ${v}^{J}_{L' S' M'B'
, \lambda_\gamma\lambda_N}(k',q)$ 
in Eq.~(\ref{eq:pw-v}), the only parameters to be determined from the
data of pion electroproduction reactions are the bare helicity amplitudes
defined above. 

As an example of the obtained results reported in Ref.~\cite{jklmss09} 
we present in Fig.~\ref{fig:5dim} a sample of the five fold differential 
cross section together with the importance of $\pi N$ intermediate 
loops in the electromagnetic production cross sections.

\section{Summary and future perspectives}

Since it was initiated in 2006, the Excited Baryon Analysis Center 
has been playing an important role in the effort to extract and 
interpret nucleon resonance properties from the extant data 
of hadronic and electromagnetic single and double meson production 
reactions. A number of publications, briefly sketched in this 
proceedings, have resulted which basically touch almost all of the 
relevant experimental data which need to be ultimately correlated 
within the same framework. The ability of the dynamical coupled 
channels model to serve as common framework to correlate hadronic 
and electromagnetic meson production reactions has been 
shown already through the several topics described in this 
proceedings. In the near future a major effort will be pursued 
to extend the description of electromagnetic reactions up to 
$W=2$ GeV and $Q^2\sim 6$ GeV$^2$, and to incorporate into the 
framework the $KY$ channels.

\begin{theacknowledgments}
I would like to thank J. Durand, 
H. Kamano, T.-S. H. Lee, A. Matsuyama,
T. Sato, B. Saghai, and N. Suzuki for the collaborations we have 
kept during the last years. I thank also R. Arndt, M. Doering, 
S. Krewald and C. Hanhart for discussions and comments on parts of 
this work, and the organizers of the Hadron 2009 conference for 
the nice meeting. This work is supported by a CPAN CSD 2007-0042 
contract, and by Grant 
No. FIS2008-1661 (Spain). The computations were performed at NERSC (LBNL) 
and Barcelona Supercomputing Center (BSC/CNS) (Spain). The authors 
thankfully acknowledge the computer resources, technical expertise 
and assistance provided by the Barcelona Supercomputing Center 
- Centro Nacional de Supercomputacion (Spain).
\end{theacknowledgments}

\bibliographystyle{aipproc} 

\end{document}